\documentclass[ALICE,manyauthors,onecolumn]{cernphprep}
\usepackage[comma,square,numbers,sort&compress]{natbib}
\usepackage{hyperref}
\usepackage{lineno}
\usepackage{siunitx}
\usepackage{wasysym}
\usepackage{placeins}
\usepackage{bm}

\usepackage[T1]{fontenc}
\usepackage{orcidlink}

\usepackage{stackengine}
\DeclareSIUnit\fm{\femto\meter}
\DeclareSIUnit\eVc{\eV\per\clight}
\DeclareSIUnit\keVc{\keV\per\clight}
\DeclareSIUnit\MeVc{\MeV\per\clight}
\DeclareSIUnit\MeVcc{\MeV\per\clight\squared}
\DeclareSIUnit\GeVc{\GeV\per\clight}
\DeclareSIUnit\GeVcc{\GeV\per\clight\squared}
\DeclareSIUnit\clight{\text{\ensuremath{c}}}
\DeclareSIUnit[number-unit-product = ]\percent{\char`\%}
\newcommand{\mycomment}[1]{}

\newcommand{\onethree}     {$\sqrt{s}~=~13$~Te\kern-.1emV\xspace}

\newcommand{\pt}{\ensuremath{p_{\rm T}}\,} 
\newcommand{\mt}{\ensuremath{m_{\rm T}}\,} 
\newcommand{\kt}{\ensuremath{k_{\rm T}}\,}

\newcommand{\kd}{\ensuremath{\mbox{K$^{+}$--d}}~}
\newcommand{\pd}{\ensuremath{\mbox{p--d}}~}

\newcommand{\piPd}{\ensuremath{\mbox{$\pi^{+}$--d}}~}
\newcommand{\piMd}{\ensuremath{\mbox{$\pi^{-}$--d}}~}
\newcommand{\piPMd}{\ensuremath{\mbox{$\pi^{\pm}$--d}}~}

\newcommand{\piPdbar}{\ensuremath{\mbox{$\pi^{+}$--$\overline{\mbox{d}}$}}~}
\newcommand{\piMdbar}{\ensuremath{\mbox{$\pi^{-}$--$\overline{\mbox{d}}$}}~}

\newcommand{\dbar}{$\overline{\mbox{d}}$}

\newcommand{\ks}           {\ensuremath{k^*}\xspace}

\newcommand{\MeVc}{\ensuremath{\mathrm{MeV}\kern-0.05em/\kern-0.02em \textit{c}}}
\newcommand{\GeVc}{\ensuremath{\mathrm{GeV}\kern-0.05em/\kern-0.02em \textit{c}}}

\newcommand{\events}{\ensuremath{1\times10^9}\,}

\newcommand{\HeThree}{$^3$He}
\newcommand{\HeFour}{$^4$He}
\newcommand{\Hype}{$^3_{\Lambda}$H}
\newcommand{\Triton}{$^3$H}
\newcolumntype{s}{>{\hsize=.5\hsize}X}

\begin{document}%

\begin{titlepage}
\PHyear{2025}
\PHnumber{081}      
\PHdate{31 March}  
%
%
\title{Observation of deuteron and antideuteron formation from resonance-decay nucleons}
\ShortTitle{Observation of deuteron and antideuteron formation from resonance-decay nucleons}

\Collaboration{ALICE Collaboration\thanks{See Appendix~\ref{app:collab} for the list of Collaboration members}}
\ShortAuthor{ALICE Collaboration} 

\begin{abstract}
High-energy hadronic collisions generate environments characterized by temperatures above 100 MeV~\cite{alicereviewarxiv,ALICE:BlastWavepp}, about 100,000 times hotter than the center of the Sun. It is therefore currently unclear how light (anti)nuclei with mass number A of a few units, such as the deuteron, \HeThree, or \HeFour, each bound by only a few MeV, can emerge from these collisions~\cite{Butler1963Jan,Kapusta1980Apr}. The ALICE collaboration reports that deuteron-pion momentum correlations in proton-proton (pp) collisions provide model-independent evidence that about 90\% of the observed (anti)deuterons are produced in nuclear reactions~\cite{Sun2024} following the decay of short-lived resonances, such as the $\Delta (1232)$. These findings, obtained by the ALICE experiment at the Large Hadron Collider (LHC) resolve a gap in our understanding of nucleosynthesis in ultra-relativistic hadronic collisions. Beyond offering insights on how (anti)nuclei are formed in hadronic collisions, the results can be employed in the modeling of the production of light and heavy nuclei in cosmic rays~\cite{PierreAuger2014} and dark matter decays~\cite{ALICE:2022zuz,Serksnyte:2022onw}. 
\end{abstract}

\end{titlepage}
\setcounter{page}{2}

The following question has long intrigued nuclear physicists: What is the microscopic mechanism behind the formation of light nuclei and antinuclei in hadron--hadron collisions~\cite{Kapusta1980Apr,Butler1963Jan}?
In ultrarelativistic heavy-ion collisions with energies per nucleon up to a few TeV ($10^{12}$ electron volts), the study of particle production is directly connected to the confinement of color charge in color-neutral hadrons. These collisions produce a quark--gluon plasma, a state in which quarks and gluons are deconfined, and as the system evolves, they bind into hadrons and light (anti)nuclei~\cite{alicereviewarxiv, SHM5}. 
Since the binding energies of light (anti)nuclei are significantly lower (2.23 MeV for deuteron, 7.72 MeV for \HeThree~\cite{Marcucci2020}) than the average kinetic energy of hadrons in such energetic collisions (order of 100 MeV~\cite{alicereviewarxiv}), the question is about both the formation of loosely-bound nuclei and their survival through the hadronic phase that follows the hadronization of the QGP. This issue is also relevant in ultrarelativistic proton--proton (pp) and proton--nucleus (p--A) collisions, where the formation of a quark--gluon plasma remains under experimental and theoretical scrutiny and average kinetic energies above 100 MeV can still be achieved~\cite{ALICE:BlastWavepp}. 

The study of (anti)nucleus formation in hadronic collisions is of critical importance in astrophysics. 
On one front, the precise composition of ultra high-energy ($>$PeV) cosmic rays, particularly their heavier-elements (A > 50) component, remains an open question~\cite{PierreAuger2014}. 
A microscopic modeling of nucleus formation in ultrarelativistic hadron collisions is an essential ingredient to understand the composition of these cosmic rays and to uncover the origin of particle acceleration mechanisms in the Universe~\cite{Petrosian:2008irtemp}. 
On another front, antinuclei formation — whether from cosmic-ray interactions with the interstellar medium or as potential products of dark-matter decay—plays a pivotal role in indirect searches for dark matter~\cite{Coalescence5,Serksnyte:2022onw,ALICE:2022zuz}.
Experimental investigations into the microscopic processes underlying light nucleus and antinucleus formation thus offer a dual benefit: they advance knowledge of the strong interaction in the non-perturbative regime and provide the quantitative framework needed to decode the spectra of cosmic rays and their origins.

The yields of nuclei such as deuterons (p--n bound system), \Triton~(p--n--n),\HeThree~(p--p--n), \HeFour~(p--p--n--n), \Hype~($\Lambda$--p--n), and their corresponding antinuclei have been precisely measured at the Relativistic Heavy Ion Collider (RHIC) in Au--Au collisions at center-of-mass energies per nucleon pair ($\sqrt{s_\mathrm{NN}}$) across an energy range from 7.7 GeV to 200 GeV~\cite{RHIC2, RHIC3,RHIC5,RHIC6} and at the LHC for pp collisions at $\sqrt{s}$ ranging from 0.9 to 13 TeV,  as well as for p--Pb and Pb--Pb collisions at $\sqrt{s_\mathrm{NN}} = 2.76-8.16$ TeV~\cite{nuclei_pp_PbPb, deuteron_pp_7TeV, deuteron_pPbALICE, 3He_pPb}.  
According to current knowledge, nuclei can be produced through either direct emission as multi-quark states following a collision, 
similar to other hadrons, such as protons or pions, or through a secondary binding mechanism of nucleons

Two types of models have been employed to study these mechanisms. \textit{Statistical hadronization models} (SHMs) describe the direct production and assume that hadrons and nuclei are directly emitted from a source in thermal and chemical equilibrium, with abundances determined by the particle mass, the system temperature, volume, and quantum number conservation~\cite{SHM2, SHM5,Donigus2022Oct}. 
This work uses the \textit{canonical statistical model} (CSM)~\cite{vanillaCSM}, which is better suited for pp collisions. While CSMs predict yields effectively, they do not provide insights into the microscopic mechanisms driving (anti)nucleus formation.
In contrast, \textit{coalescence models}~\cite{Coalescence2,Coalescence5, Coalescence7,Mrowczynski:1987oid,Mrowczynski:2019yrr} emulate binding mechanisms and they assume that (anti)nucleons form independently before binding to create (anti)nuclei. This approach incorporates microscopic parameters, such as the spatial proximity of nucleons alongside their strong interactions, allowing for a satisfactory description of yields 
and momentum distributions~\cite{Coalescence7,PHQMD1}. 

Microscopic calculations implemented in event generators for heavy-ion collisions~\cite{Neidig:2021bal,Sun2024} include pion-catalyzed reactions — both formation and disintegration (e.g. $\pi +p +n \iff \pi +d$) —  and successfully describe measured nuclear yields. The important aspect of such models is that a third body, such as a meson, aids the binding process by carrying away the excess energy.

Overall a direct experimental evidence for the microscopic mechanisms of (anti)nucleus formation remains absent. Femtoscopy provides a complementary approach by examining pion-(anti)deuteron ($\pi$--d) momentum correlation and offers direct insights into the microscopic processes underlying (anti)deuteron formation.
This technique has been effectively employed by the ALICE Collaboration to study various hadron pairs produced in pp and p--Pb collisions at the LHC, see e.g.~\cite{PhysRevX.14.031051} and the references therein, 
shedding light on their residual strong interactions. 

Using $\pi$--d femtoscopy correlations, the study presented in this paper reveals, in a model-independent manner, that (anti)deuterons are formed following the decay of strong resonances, such as the $\Delta$(1232) (hereafter "$\Delta$"). Considering the possible contribution to all produced resonances, we estimate that 88.9 $\pm$ 6.3\% of the observed (anti)deuterons are generated through binding processes.
These findings resolve a longstanding puzzle regarding the formation of light (anti)nuclei in collider experiments and provide a robust foundation for further modeling of nucleosynthesis from hadronic collisions, both in accelerators and in the Universe.

\section*{Resonances and correlation function}

As a bound state of a proton and a neutron, the deuteron may inherit correlations developed by its constituent nucleons during the evolution of the collision. The study of pion--deuteron ($\pi$--d) correlations offers a sensitive probe of this process. If deuterons are produced thermally alongside pions, any correlation signal between them would arise solely from final-state interactions. In contrast, if deuterons form through the coalescence of nucleons originating from the decay of an intermediate resonance, the resulting $\pi$--d correlation function could retain the signatures of that resonance. 

The correlation function $C(k^*)$ is the key experimental observable, and $k^*$ is the single-particle momentum in the pair rest frame (PRF). 
Experimentally, $C(k^*) = \mathcal{N}\left[N_\text{same}(k^*)/N_\text{mixed}(k^*)\right]$, where $N_\text{same}(k^*)$ (same-event sample) is the distribution of relative momenta between the $\pi$--d pair measured for pions and deuterons stemming from the same collision~\cite{ALICE:2018ysd}. Equivalently, $N_\text{mixed}(k^*)$ (mixed-event sample) is an uncorrelated reference obtained by building the distribution via the combination of pions and deuterons originating from different collisions. Lastly, $\mathcal{N}$ is a normalization factor ensuring the proper convergence of $C(k^*)$ to unity at large $k^*$. Indeed, in the case of non-interacting particles, the correlation function is equal to unity for all $k^*$ as the relative momentum distribution is purely governed by the underlying single-particle phase space, which is the same for $N_\text{same}(k^*)$ and $N_\text{mixed}(k^*)$ distributions. An attractive interaction enhances the correlation function above unity at low $k^*\lesssim$ 200~MeV/$c$, while a repulsive interaction leads to a depletion below unity. A resonance which decays into the hadron pair of interest would produce a peak in the $k^*$ spectrum.

\begin{figure}[!hbt]
    \centering
    \includegraphics[width=1.0\linewidth]{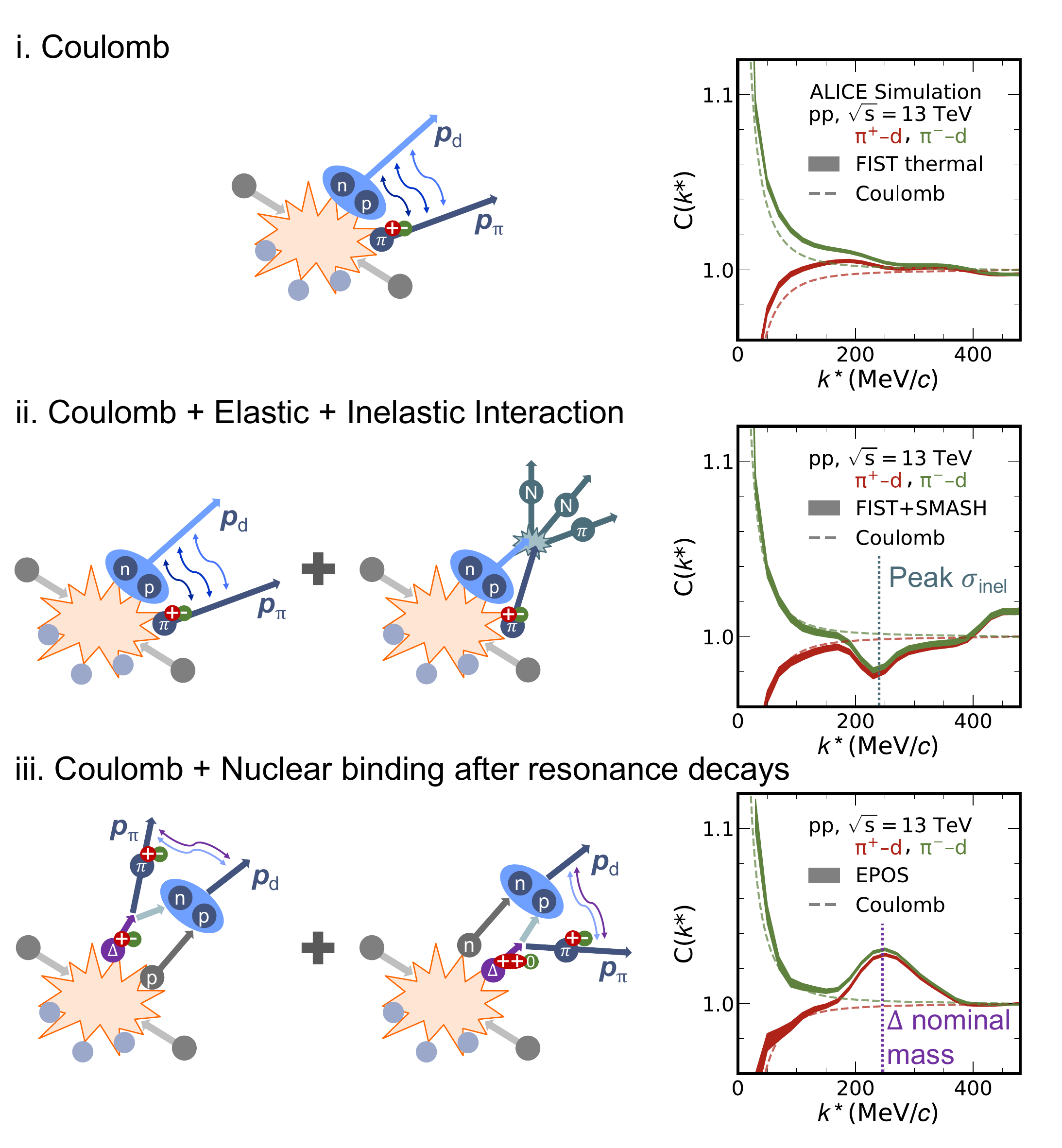}
    \caption{\textbf{(Anti)deuteron production scenarios.} Illustration of three scenarios for deuteron production and interaction with pions (left) and the resulting \piPMd correlation functions (right). All scenarios include Coulomb attraction between $\pi^-$--d (green curves) and Coulomb repulsion between the $\pi^+$--d (red curves). The dashed lines always show the correlation function using Coulomb interaction. Scenarios (i) and (ii) represent thermally-produced deuterons with only Coulomb and Coulomb+elastic+inelastic interactions, respectively. Scenario (iii) depicts deuteron formation via nuclear binding following $\Delta$-resonance decays. All the simulations include the charge conjugates ($\piPd \equiv \piPd\oplus \piMdbar$ and  $\piMd  \equiv \piMd\oplus\piPdbar$). The bands width corresponds to the statistical uncertainties of the models.}
    \label{fig:sketch}
\end{figure}

Figure~\ref{fig:sketch} illustrates three scenarios of (anti)deuteron production mechanisms and interactions and the resulting $\pi^\pm$--d correlations.
Simulations were performed to quantify the effects of the different production scenarios, and details are provided in \nameref{sec:Methods}.
All scenarios include repulsive (red curves) and attractive (green curves) Coulomb interactions for the $\pi^+$--d and $\pi^-$--d systems, respectively. The strong interaction contribution is minimal due to the small scattering parameters of the \piPMd system and is hence neglected~\cite{Meissner:2005bz, Hauser1998}. 
In scenarios (i) and (ii), directly-produced deuterons and pions are considered. The simulation results shown on the right panel were obtained assuming that pions and (anti)deuterons are produced following a canonical statistical hadronization scheme, ThermalFIST~\cite{TheFIST}. The obtained correlation functions are multiplied by the Coulomb correlation function. The results display a depletion (enhancement) in the correlation function at low \ks for $\pi^+$--d ($\pi^-$--d) due to the Coulomb interaction.
 
In scenario (ii), the elastic and inelastic scattering of pions and deuterons is considered. This is tested by using the hadronic transport model SMASH~\cite{SMASH} as an afterburner to ThermalFIST to simulate inelastic and elastic rescattering according to the experimental cross sections. The elastic processes do not modify the shape of the correlation function, as both the incoming and outgoing $\pi^\pm$--d pairs must conserve energy, ensuring that their relative momentum \ks remains unchanged. The same holds for pseudo-elastic processes, in which an intermediate $\Delta$ resonance is formed, as in $\pi+(\text{pn}) \rightarrow \text{p}\Delta \rightarrow \pi+(\text{pn})$.

Inelastic $\pi^\pm$--d scattering, on the other hand, leads to deuteron destruction, reducing the number of measurable pairs in the \ks region where the inelastic cross section reaches its maximum. Both elastic and inelastic cross sections peak at the nominal $\Delta$ mass (\ks$\approx240$ MeV/$c$), and the inelastic one is three times larger than the elastic contribution~\cite{Coci2023Jul}.
The right panel of scenario (ii) in Fig.~\ref{fig:sketch} shows the results of these simulations for the \piPd and \piMd cases. As expected, a depletion at the relative momentum \ks$\approx240$ MeV/$c$, corresponding to the peak of the inelastic cross section, is observed.

In scenario (iii), a deuteron forms when a primordial nucleon binds with one from a 
$\Delta$ decay. These resonances are very short-lived excited states of nucleons and decay after approximately 1.5 fm/$c$ into $\pi$--nucleon pairs. Considering all charge states ($\Delta^{++,\,+,\,0,\,-}$), they are expected to contribute $43\%$ of the nucleon yield in pp collisions at the LHC~\cite{Vovchenko2017Jan,TheFIST}. Measurements of the $\pi^{\pm}$--p femtoscopy correlations by ALICE~\cite{ALICE:pPi} have already shown the presence of the $\Delta$ resonances, modified by rescattering and regeneration effects (see \nameref{sec:Methods}). For the formation of deuterons, the possible combinations include neutron-proton binding from $\Delta^{++}\rightarrow \pi^+$--p 
 or $\Delta^0 \rightarrow \pi^-$--p, proton-neutron binding from $\Delta^\pm \rightarrow \pi^\pm$--n, and binding of two nucleons from separate $\Delta$ decays.
 This scenario was simulated by exploiting a state-of-the-art coalescence afterburner~\cite{Coalescence7} combined with the EPOS 3 event generator~\cite{Werner:2010aa,Werner:2013tya}. The latter accounts for resonance production and their decays, while the aforementioned rescattering and regeneration effects are not included in the simulations. The results are shown in the right panel, and a clear peak appears in correspondence with the $\Delta$ resonance nominal mass. The observed peak is due to the residual correlation between the pion and an (anti)nucleon from the $\Delta$ decay during the (anti)deuteron formation process. 

The three patterns in the correlation function correspond to different physics scenarios and are easily distinguishable from each other.
 The observed patterns remain unchanged across different models: while model parametrization may shift or rescale the structures, their shapes are preserved. This constitutes a solid reference to proceed with the interpretation of the experimental data.

\section*{Results and discussion}
\label{results}

\begin{figure}[t]
    \centering
    \subfigure{\centering \includegraphics[width=0.49\linewidth]{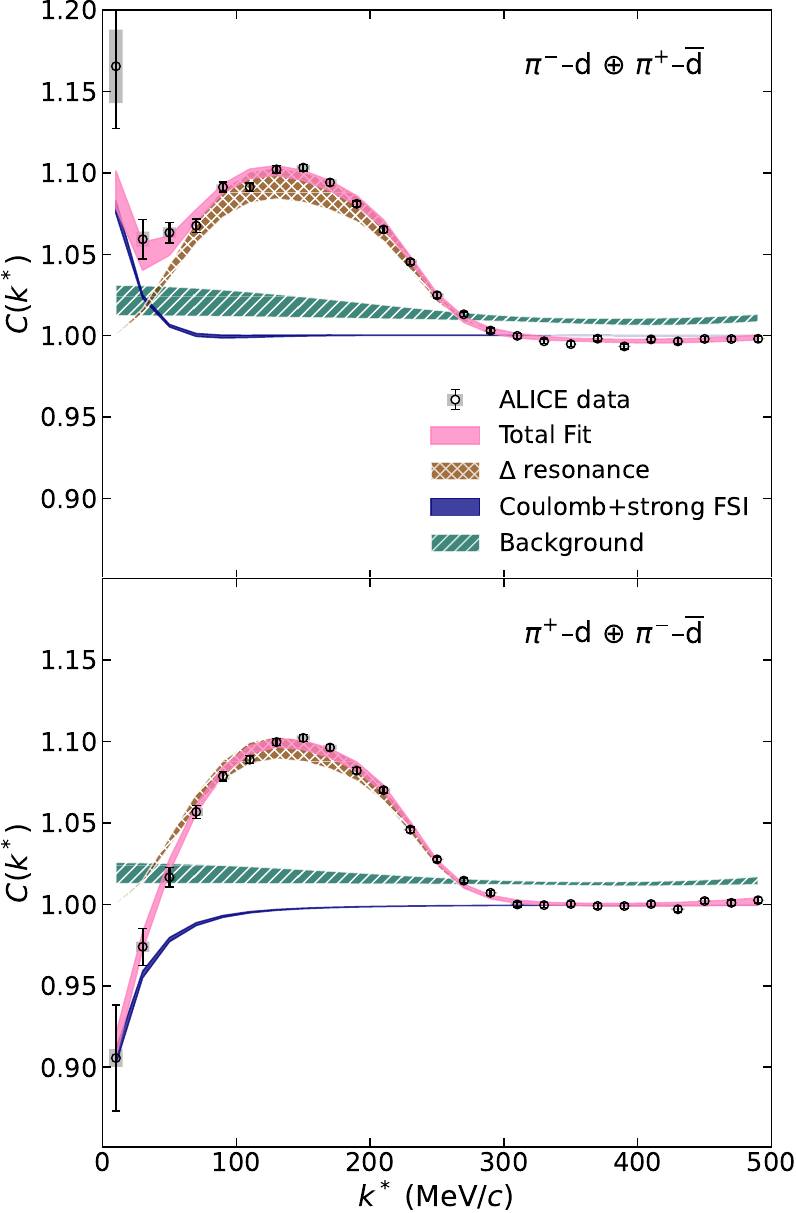}}
    \caption{\textbf{Experimental \piMd and \piPd correlation functions.} The data are obtained from high-multiplicity pp collisions at $\sqrt{s} = 13$~TeV. The top panel shows the measured \piMd correlation function together with the corresponding fit function (magenta). The brown cross-hatched band represents contributions from the $\Delta$ resonance, the blue band denotes the Coulomb and strong FSI interactions, and the teal diagonally hatched band corresponds to the residual background. The widths of the bands indicate the fit uncertainty. The lower panel depicts, in the same representation, the \piPd correlation function. However, the strong FSI interaction is neglected for this system. The $\chi^2$ per degree of freedom is 14/15 for both correlations.}
    \label{fig:pi-dfits}
\end{figure}

The experimental \piPd and \piMd correlation functions have been measured in pp collisions at $\sqrt{s}=13$~TeV. Charged pions ($\pi^{\pm}$), deuteron (d), and antideuteron (\dbar) tracks are reconstructed with the ALICE detector, and their momentum transverse to the beam direction (\pt) is measured in the range \pt~$\in$ $[0.14,4.0]$~GeV/$c$ for pions and $\pt \in [0.8,2.4]$~GeV/$c$ for deuterons. The excellent particle identification and tracking capabilities of the ALICE detector provide samples of $\pi^{+} (\pi^{-})$, d (\dbar) with a purity of 99\% and 100\%, respectively. Further details on the particle selection and evaluation of the systematic uncertainties are described in \nameref{sec:Methods}. After the selection of pions and (anti)deuterons, the correlation functions for pairs of particles (\piPd and \piMd) and their charge conjugates (\piPdbar and \piMdbar) are obtained. Since the same interaction governs hadron--hadron and antihadron--antihadron pairs~\cite{ALICE:Run1}, the sum of particles and antiparticles is considered ($\piPd \equiv \piPd\oplus \piMdbar$ and  $\piMd  \equiv \piMd\oplus\piPdbar$) in the following. The resulting \piMd and \piPd correlation functions are depicted by the open markers in the top and bottom panels of Fig.~\ref{fig:pi-dfits}. The gray boxes around the markers represent the systematic uncertainties, while the vertical bars show the statistical uncertainties. The fit results for the \piMd and \piPd correlation functions are displayed in the top and bottom panels, respectively.  

The measured \piPMd correlation functions are modeled and fitted using a decomposition approach summarized by the relation 
$C_\text{fit}(k^*) = \epsilon(k^*) \otimes B(k^*) \left[\lambda_\text{gen} C_\text{gen}(k^*) + (1 - \lambda_\text{gen})\right]$ (see details in \nameref{sec:Methods}).
Here, $\epsilon(k^*)$ represents a correction for momentum resolution effects, and $B(k^*)$ is a baseline accounting for residual background correlations. The latter arise mainly from non-primary pions produced in weak decays of long-lived resonances, as well as from secondary particles originating from interactions with the detector material. Such contributions can mimic correlated pairs and must be accounted for in the modeling. The parameter $\lambda_\text{gen}$ quantifies the fraction of genuine \piPMd pairs, with the non-genuine component primarily arising from the feed-down of long-lived resonances into pions~\cite{ALICE:2023sjd} with a life-time $\tau > 5 \,\text{fm}/c$. The term $C_\text{gen}(k^*)$ denotes the corresponding genuine correlation function that contains Coulomb and strong interactions, alongside contributions from the $\Delta$ resonance. The interaction components are modeled using the CATS (Correlation Analysis Tool using the Schrödinger equation) framework \cite{Mihaylov:2018rva}. 
Theoretically, $C(k^*)=\int d^3 r^* S(r^*)\times \left|\psi(\boldsymbol{k}^*,\boldsymbol{r}^*)\right|^2$, where $r^*$ is the relative distance (in the PRF) between the particles at the time of their effective emission, $\psi(\boldsymbol{k}^*,\boldsymbol{r}^*)$ is the wavefunction of the pair relative motion, and $S(r^*)$ is the source function corresponding to the probability to emit the pair at a certain relative distance $r^*$~\cite{Lisa:2005dd}. Dedicated studies of the source function in pp collisions at $\sqrt{s}=13$~TeV performed by the ALICE Collaboration revealed a common emission source for all hadrons~\cite{Acharya2025, ALICE:2023sjd, ALICE:pPi}. This source is typically modeled by a Gaussian function with a standard deviation (an effective size of the source) of $r_\text{eff} \approx1.5$~fm, obtained by accounting for the contribution of short-lived resonances (see \nameref{sec:Methods} for details).

For the source, an effective Gaussian distribution with $r_\text{eff} = 1.51\pm0.12$~fm was employed (details in \nameref{sec:Methods}). 
The real part of the \piMd potential is included in the fit, however due to the small scattering parameters of the \piPMd system, the contribution is negligible~\cite{Hauser1998,Meissner:2005bz}.
 In order to gauge the influence of the resonance decays on the \piPMd correlations, the contributions of the $\Delta$ resonances extracted from the measured $\pi^{\pm}$--p correlations are modified assuming that the nucleon emerging from the $\Delta$ decay coalesces with an additional nucleon to form a deuteron. The assumption is that the two nucleons have similar momenta (see \nameref{sec:Methods} for details). Finally, the relative momentum \ks between the pion from the $\Delta$ decay and the deuteron is evaluated. All charge states $\Delta^{++,\,+,\,0,\,-}$ are considered, assuming the $\Delta^{+,\,-}$ peak has the same shape as $\Delta^{++,\,0}$ from $\pi^{\pm}$--p correlations. The experimental correlation functions are well described, confirming the scenario where the deuteron is formed after the decay of the $\Delta$ via a fusion process as assumed in scenario (iii) in Fig.~\ref{fig:sketch}. An excellent description (see Fig.~\ref{fig:pi-dfits}) of the measured correlation function is obtained by adopting the data-driven shape of the $\Delta$, derived from $\pi^\pm$--p correlations~\cite{ALICE:pPi}.  
 The $\Delta$ shape in the \piPMd correlation function exhibits a shift toward lower masses due to rescattering effects, consistent with the displacement observed in Fig.~\ref{fig:pi-dfits} relative to the nominal $\Delta$ position at \ks$\approx$ 240 MeV/$c$. 
 The simulations obtained with the EPOS 3 event generator currently do not include the rescattering of the $\Delta$ decay products, resulting in no shift of the $\Delta$ peak in scenario iii) of Fig.~\ref{fig:sketch}.

The evidence of $\Delta$ decay in the \piPMd correlation function is model-independent, as freeing the radius parameter does not affect the results, the Coulomb interaction is inherently model-independent and the residual background 
is accounted for in the fit's systematic uncertainties.

Furthermore, the fraction of deuterons produced following a resonance decay is extracted. The contribution from $\Delta$ resonances is evaluated by integrating the peak in the \piPMd correlation functions corresponding to these resonances, subtracting the number of \piPMd pairs expected in the same $k^*$ region without a nucleon originating from the $\Delta$ resonance (Coulomb + background), and dividing the result by the total number of detected deuterons. The result is corrected for combinatorial effects, reconstruction efficiency, and the non-measured $\pi^0$ final state. With these corrections, the fraction of deuterons produced via a $\Delta$ resonance is calculated to be $60.6 \pm 4.1~\%$ (details in \nameref{sec:Methods}). 
To test the compatibility of this measurement with expectations from event generators, the EPOS 3 model is used. For this, the yield of baryonic resonances in EPOS 3 was adjusted to match the predictions of the CSM Thermal FIST for pp collisions at $\sqrt{s}=13$~TeV. Furthermore, the change in the pion detector acceptance resulting from the shift in the experimentally observed $\Delta$ spectral shape is taken into account. In this simulation, the fraction of deuteron where at least one of the nucleons stems from any $\Delta$ resonance and both the deuteron and the $\pi$ are found within the acceptance is determined to be $53.8\pm3.1~\%$ (more details in \nameref{sec:Methods}).
The fact that these two fractions are in agreement within $1.32$ standard deviations demonstrates that the survival probability of deuterons produced in pp collisions at the LHC is very high. This is the case since (anti)nuclei are produced after the resonance decays, where spectral temperatures of about 20 MeV (see \nameref{sec:Methods} for details) have been evaluated, much lower than the average kinetic energy of hadrons ($\sim$ 100 MeV) in pp collisions at the LHC. 

Deuterons can form through fusion after any strong resonance decay, with $\Delta$ resonances accounting for 77.3 $\pm$ 1.2 \% of all cases. In addition, 
considering that 15.5 $\pm$ 0.5\% of $\Delta$ resonances are not properly reconstructed due to the detector acceptance effects, the experimental fraction of deuterons from $\Delta$ can be scaled up to a total fraction of deuterons originating from all resonances of 88.9$\pm$ 6.3\%.
These results demonstrate not only that the presence of resonances contributes to the (anti)deuteron production but also that it is the dominant process responsible for the creation of deuterons.

\section*{Summary}

In this work, \piPMd correlation functions measured in pp collisions at $\sqrt{s}=13$~TeV by the ALICE collaboration at the LHC are used to study the (anti)deuteron production mechanism. It is demonstrated for the first time that (anti)deuteron formation via nucleonic fusion follows the strong decay of short-lived resonances. Model-independent evidence is provided by observing the residual correlation of pion-nucleon pairs stemming from the same $\Delta$ decay in the pion-deuteron correlation function. This effect can only be explained assuming that (anti)deuteron formation occurs after the $\Delta$ decay and the measured correlation is interpreted via a data-driven method based on the independent measurement of the  $\Delta$ in the $\pi^{\pm}$--p final state. The residual signal in the \piPMd correlations can be used to evaluate the fraction of (anti)deuterons produced following $\Delta$ decays, which is found to be $60.6 \pm 4.1~\%$. Extending this reasoning to all strong resonances produced in pp collisions at $\sqrt{s}=13$~TeV it is found that 88.9 $\pm$ 6.3\% of (anti)deuterons are formed through binding processes involving nucleons originating from strongly decaying resonances.
This large fraction demonstrates that most of the (anti)nuclei are produced 
through secondary binding processes in pp collisions at the LHC and not by direct emission as other hadrons. A large survival probability is expected for (anti)deuterons since the low spectral temperature of $\Delta$ ($\sim$20 MeV) reflects that the environment where (anti)nuclei are created is characterized by a much lower kinetic energies than the hadronization phase ($\sim$100 MeV) in pp collisions at the LHC. These findings solve a longstanding puzzle in nuclear physics, providing insight into the microscopic mechanism that leads to (anti)nuclei formation in pp collisions at the LHC. These insights can now be employed for a more realistic microscopic modeling of (anti)nuclei production, e.g. in cosmic ray induced reactions. 


\bibliographystyle{utphys}   
\bibliography{Content/Bibliography.bib}

\newpage
\appendix

\section*{Methods}
\label{sec:Methods}
\subsection*{Event selection}
The results are based on the analysis of a dataset comprising inelastic pp collisions at $\sqrt{s}=13$~TeV, recorded with the ALICE detector~\cite{Aamodt:2008zz, Abelev:2014ffa} during the LHC Run 2 (2015--2018). The events are selected using a high-multiplicity (HM) trigger, which captures the highest multiplicity events -- specifically, the top 0.17\% of all inelastic collisions that include at least one charged particle within the pseudorapidity interval $|\eta| < 1$ (denoted as 0.17\% INEL $>$ 0). This approach ensures a statistically rich sample, since a five-fold increase in the production of (anti)deuteron candidates has been observed in HM pp collisions compared to minimum bias pp collisions~\cite{nuclei_pp_13TeV_HM}. 
The sample of HM triggered collisions considered for this analysis corresponds to \events events. On average, 31 charged tracks are found within $|\eta| < 0.5$~\cite{ALICE:2020mfd} for the HM-triggered collisions. Detailed descriptions of the event selection criteria, pileup rejection techniques, primary-vertex reconstruction methods, and the HM trigger procedure are provided in Ref.~\cite{Acharya:2019kqn}.

\subsection*{Tracking and particle identification}
Particle identification and momentum measurement of charged particles are performed using the Inner Tracking System (ITS)~\cite{Aamodt:2010aa}, Time Projection Chamber (TPC)~\cite{Alme:2010ke}, and Time-of-Flight (TOF)~\cite{Akindinov:2013tea} detectors of ALICE  covering the whole azimuthal angle and the pseudorapidity interval $|\eta|<\num{0.9}$. These detectors are located within a uniform magnetic field of \SI{0.5}{T} along the beam axis, generated by the ALICE solenoid magnet, which causes the trajectories of particles to bend. The curvature of the charged-particle tracks is used to measure the particle momenta. The transverse momentum for pion and deuteron candidates is determined with a resolution ranging from approximately \SI{2}{\percent} for tracks with $\pt \approx \SI{10}{}$~GeV/$c$ to below \SI{1}{\percent} for $\pt<\SI{1}{}$~GeV/$c$. Particle identification is performed by measuring the energy loss per unit track length (\(\mathrm{d}E/\mathrm{d}x\)) in the TPC detector and the particle velocity (\(\beta\)) in the TOF detector. For tracks in the TPC detector, the signal is obtained from the \(n\sigma_{\text{TPC}}\) distribution, where \(n\sigma_{\text{TPC}}\) represents the deviation of the measured signal from the expected value for a given particle hypothesis, normalized by the detector resolution. Similarly, for the TOF detector, the resolution is defined by \(n\sigma_{\text{TOF}}\), which quantifies the difference between the measured and expected time of flight, also normalized by the resolution. Additional experimental details are discussed in Ref.~\cite{Abelev:2014ffa}. The selection criteria for pion and deuteron tracks used in this work are described in Refs.~\cite{PhysRevX.14.031051, ALICE:2023sjd}.

Pions are identified via the measurement of the specific energy loss within $|n\sigma_{\text{TPC}}|<3$ in a transverse momentum range $ p_{\mathrm{T}} \in [0.14, 4.0]$~\GeVc. This information is combined with the time-of-flight measurement by taking the geometric sum, $ \sqrt{n\sigma_{\text{TPC}}^2 + n\sigma_{\text{TOF}}^2} < 3 $, for track momentum $ p > 0.5$~\GeVc. Similarly, the deuteron candidates are selected within a transverse momentum range $ p_{\mathrm{T}} \in [0.8, 2.4]$~\GeVc. They are identified by employing $| n\sigma_{\text{TPC}}|< 3$ for candidate tracks with momentum $ p < \SI{1.4}{}$~\GeVc, while both TPC and TOF information are required, $\sqrt{n\sigma_{\text{TPC}}^2 + n\sigma_{\text{TOF}}^2} < 3 $, for candidates with $p > \SI{1.4}{}$~\GeVc. Additionally, for (anti)deuteron candidate selections, electrons are rejected by the condition $n_{\sigma,\mathrm{TPC},\mathrm{e}}> 6 $ for $ p < 1.4$~GeV/$c$ and pions are rejected by the condition $ n\sigma_{\mathrm{TPC},\pi} > 3$ for the tracks with momentum $ p > 1.4$~\GeVc. Overall, using these methods, a purity of  99\% for $\pi^\pm$ and 100\% for (anti)deuterons is achieved.

 The selection criteria of pions and deuterons constitute the primary source of systematic uncertainties associated with the measured correlation function. All particle selection criteria are varied from their default values. To account for the effect of possible correlations, the analysis of $\piPd$ ($\piMd$) pairs is repeated $44$ times using random combinations of such selection criteria. 
 The total systematic uncertainties are extracted by first randomly selecting a correlation function from the 44 systematic variations. For each sampled function, a bootstrap method is applied by randomly varying the $C(k^*)$ values in the individual $k^*$ bins according to their statistical uncertainties, assuming Gaussian errors. This results in a distribution of values for each $k^*$ bin, which is then fitted to determine the total uncertainty. Since the statistical and systematic uncertainties are independent, the total uncertainty is obtained by adding them in quadrature. The systematic component is then determined by subtracting the known statistical uncertainty. The systematic uncertainties are largest at low \ks $\approx 10$ MeV/$c$, reaching \SI{1}{\percent}. The same procedure is applied to extract the uncertainties of the fitted parameters and propagated to the final results on the fraction of deuterons stemming from resonance-assisted fusion processes.
 
\subsection*{Characterization of the particle-emitting source}
A standard approach to evaluate the source function, used by ALICE in pp collisions, is the Resonance Source Model (RSM)~\cite{Acharya2025, ALICE:2023sjd}. Within these publications, the ALICE Collaboration measured the source size for baryon–baryon, meson–baryon, and meson–meson pairs, demonstrating a common emission source of all particles and resonances produced directly in the collision. These are described as primordial particles, while the short-lived resonances that decay into the pairs of interest on the timescale of fm$/c$ will lead to an increase of the effective source size. If this increase of the source size is properly modeled by a Monte Carlo simulations, the underlying primordial source has a Gaussian profile of width $r_\text{core}$, and scales as a function of the pair transverse mass $\smash{\mt = \left(\kt^{2} + m^{2}\right)^{1/2}}$, where $m$ is the average mass, the average of the masses of the two particles constituting the pair and $\smash[b]{\kt =\,\,\mid \bm{p}\bf{_{\textrm{T},\textrm{1}}} + \bm{p}\bf{_{\textrm{T},\textrm{2}}}\mid}$/2 is the average transverse momentum of the pair~\cite{Acharya2025, ALICE:2023sjd}. The scaling of the primordial source size follows a power law $r_\text{core} = a \left<\mt\right>^b + c$, where the parameters for the high-multiplicity pp collisions at $\sqrt{s}=13$~TeV used for the present $\pi$--d analysis are provided in Ref.~\cite{Acharya2025}. The knowledge of both the pair average $\mt$ and the cocktail of contributing resonances allows us to evaluate both the $r_\text{core}$ and subsequently the total source distribution $S(r^*)$. The present analysis incorporates the resonances decaying into pions from the ThermalFIST model~\cite{Vovchenko2017Jan,TheFIST}, as already performed in the ALICE $\pi$--$\pi$ and p--$\pi$ analyses~\cite{ALICE:2023sjd, ALICE:pPi}. 
From the study of \pd and \kd correlations in pp collisions at \onethree~\cite{PhysRevX.14.031051}, it has been shown that in pp collisions, the hadron--deuteron pairs follow the same transverse mass scaling as other hadron--hadron pairs, allowing to constrain the $\pi$--d emission source using the RSM. 
The deuterons are not produced directly by resonances. Nevertheless the present work demonstrates that resonances decaying into nucleons are an important step of the production mechanism. This will lead to an effective delay of the deuteron production, an effect already described in a previous analysis of the $K^+$--d analysis~\cite{VazquezDoce:2024nye}. The present analysis adopts a conservative approach and integrates two extreme scenarios for the deuteron production as part of the systematic uncertainties, namely assuming either that all deuterons are primordial or that the deuteron formation is delayed based on the amount of emission delay by which their constituent nucleons are affected~\cite{Acharya2025}. This variation, which affects the effective source size, $r_\text{eff}$ of up to 0.08~fm, is included in the systematic uncertainties on the modeling of the correlation functions. The final values for the $r_\text{eff}$, after the inclusion of resonances, are summarized in Table~\ref{tab:TabmValuesSourceLambda} along with the total uncertainties.

\subsection*{Corrections of the Correlation Function}
The experimental correlation function, defined as $C(k^*) = \mathcal{N}\left[N_\text{same}(k^*)/N_\text{mixed}(k^*)\right]$ is only corrected by a normalization constant $\mathcal{N}$, by ensuring that the correlation becomes unity for $k^*\in(400,~600)$~MeV$/c$. The remaining corrections are included in the fit function 
\begin{equation}\label{eq:fitfun}
C_\text{fit}(k^*) = \epsilon(k^*) \otimes B(k^*) \left[\lambda_\text{gen} C_\text{gen}(k^*) + (1 - \lambda_\text{gen})\right]. 
\end{equation}
The parameter $\epsilon(k^*)$ incorporates momentum resolution effects, which are included by obtaining a transformation matrix that can be used to apply resolution effects to the correlation functions. Details on the procedure are provided in the supplemental materials to~\cite{Acharya2025}. The required experimental inputs are the matrix itself and the experimental mixed event sample, both of which are provided in the HEPData entry related to this work. The baseline $B(k^*) = a + b{k^*}^2+c{k^*}^3$ accounts for any remaining long-range correlations~\cite{ALICE:2021njx}. These correlations do not contribute as an additive contamination to the correlations as misidentified particles do, but rather stem from the kinematics of the collision event. These long-range correlations are not correlated to the final-state interaction and can therefore be factorized and included as a multiplicative factor in the correlation.
All of the parameters of the baseline are left free within the fit procedure. The final correction to the correlation function is $\lambda_\text{gen}$, which represents the amount of genuine $\pi$--d pairs. In the context of the source, a genuine particle is either a primordial or the decay product of a short-lived resonance of lifetime $c\tau<5$~fm/$c$. Details on the extraction of these parameters for the pions and deuterons are provided in~\cite{ALICE:2023sjd} and~\cite{PhysRevX.14.031051}, respectively. Combining the information for the two species, the correction obtained for \piPMd is summarized in Table~\ref{tab:TabmValuesSourceLambda}. 
The $(1 - \lambda_\text{gen})$ factor in the definition of $C_\text{fit}(k^*)$ reflects the remaining non-genuine correlations, which are assumed to produce a flat correlation signal. These non-genuine correlations stem from misidentified particles, as well as feed-down from long-lived resonances. Due to the high purity within the present analysis, the non-genuine correlations are predominantly linked to the feed-down into pions from non-strong decays, such as decays of kaons~\cite{ALICE:2023sjd}. There is no contribution to the non-genuine correlation from feed-down into deuterons, as such decay processes do not exist, except for the weak decay of the hypertriton $({}_{\Lambda}^{3}\text{H}\rightarrow\pi^{-}+\mathrm{p}+\mathrm{d})$, which has a negligible effect.

\subsection*{$\Delta$ spectral shape}

 In the measurement of the $\pi^{\pm}$--p correlation functions~\cite{ALICE:pPi}, a prominent peak around $k^* = 211$~MeV/$c$ can be seen, associated with the $\Delta$ resonances ($\Delta^{++}$ for $\pi^+$--p and $\Delta^0$ for $\pi^-$--p). 
  In the decay of a $\Delta$ resonance into a pion–nucleon pair, the $\Delta$ is at rest in the center-of-mass frame of the decay products. By applying energy and momentum conservation for this two-body decay, the invariant mass of the $\Delta$ is related to the relative momentum $k^*$ of the decay products via
\begin{equation}
m_\Delta = \sqrt{(k^*)^2 + m_\pi^2} + \sqrt{(k^*)^2 + m_\text{N}^2}\,. \label{Eq. Delta Mass k*}
\end{equation}
Inverting this expression yields the expected relative momentum $k^*$ associated with a given $\Delta$ mass
\begin{equation}
    k^* = \frac{\sqrt{ (m_\Delta^2 - m_\pi^2 - m_\text{N}^2)^2 - 4m_\pi^2 m_\text{N}^2}}{2m_\Delta}\,. \label{Eq. kStar Mass Reso relation}
\end{equation}
For the nominal $\Delta$ mass of $M_\Delta = 1.215~\text{GeV}/c^2$, this corresponds to a relative momentum of $k^* = 211~\text{MeV}/c$ for the pion–nucleon pair. The peak position observed in the $\pi^{\pm}$--p correlations is shifted to lower values than the nominal $\Delta$ mass due to the rescattering of the decay products and regeneration of the resonances~\cite{STAR:2008twt, Reichert:2019lny, Reichert:2022uha}.

Following Ref.~\cite{ALICE:pPi}, the $\Delta$ spectral shape is modeled as  $C_{\Delta}(k^*) = \mathcal{N}_{\Delta} \times PS(p_{\text{T},\Delta}, T) \times \text{Sill}(M_{\Delta}, \Gamma_{\Delta})$. The first term $\mathcal{N}_{\Delta}$ is a normalization constant, while the undisturbed spectral shape of the resonances is described via the Sill distribution~\cite{GiacosaSill}, which depends on the resonance mass $M_{\Delta}$ and width $\Gamma_{\Delta}$.  Since the Sill is expressed as a function of $k^*$, it is essential to account for the Jacobian factor $\left|\text{d}m_\Delta/\text{d}k^*\right|$ in the change of variables, where $m_\Delta$ is given by Eq.~\eqref{Eq. Delta Mass k*}. Modifications of the spectral shape due to rescattering and resonance regeneration effects are incorporated via a multiplicative  $PS(p_{\text{T},\Delta},T)$ term~\cite{ Reichert:2019lny, Reichert:2022uha}, a Boltzmann-like phase space factor,
\begin{equation}\label{Eq. PS Term}
    PS(p_{\text{T},\Delta}, T) \propto \frac{M}{\sqrt{M^2+p_{\text{T},\Delta}^2}}\exp\left[ - \frac{\sqrt{M^2+p_{\text{T},\Delta}^2}}{T} \right]\,, 
\end{equation}
acting as a weight for the emission of the resonance with certain transverse momentum $p_{\text{T},\Delta}$ at a temperature $T$. The latter is referred to as the ``$\Delta$ spectral temperature'' in Ref.~\cite{ALICE:pPi}.

To obtain the corresponding spectral shape within the $\pi$--d correlation, a simple approach is adopted by assuming that each measured deuteron consists of two nucleons of equal momenta. In this way, the $\Delta$ spectral shape in the $\pi$--N system (Eq.~\eqref{Eq. kStar Mass Reso relation}) can be transformed into the $\pi$--d PRF. For a nominal $\Delta$ mass of $M_\Delta = 1.215~\text{GeV}/c^2$ leads to $k^* = 237~\text{MeV}/c$  in the pion-deuteron system.  

A final systematic check was performed by allowing a non-zero relative momentum between the two nucleons forming a deuteron. For this, a relative momentum sampled from a distribution, which was obtained from a coalescence model~\cite{Coalescence8}, was used. The relative momentum is, on average, $\simeq100$ MeV/$c$~\cite{Coalescence8}. The final shape of the $\Delta$ peak in the $\pi$--d correlation remains identical regardless of the assumption of the relative momenta between the nucleons. Thus, the simpler approach of identical nucleon momenta was used in the analysis. 

\subsection*{Fitting the $\pi$--d correlation}

The fit function is defined by Eq.~\ref{eq:fitfun}. The genuine correlation $C_\text{gen}(k^*)$ encapsulates Coulomb and strong interactions alongside contributions from the $\Delta$ resonance. The interaction components were modeled using the CATS framework~\cite{Mihaylov:2018rva}, which employs the Schr\"odinger equation and requires as input the source function and the strong interaction potential. The contribution of the strong interaction is minimal due to the small scattering parameters of the $\pi$--d system, since the scattering length is canceled for $\pi$--p and $\pi$--n pairs~\cite{Meissner:2005bz, Hauser1998}.
The real part of the $\pi^-$--d potential was included in the fit~\cite{Hauser1998, Meissner:2005bz}. 
To account for the $\Delta$ resonance, a phenomenological approach was adopted, expressing the genuine correlation as 
\begin{equation}\label{eq:fit_def}
C_\text{gen}(k^*) = C_\text{interaction}(k^*) \left[F_\Delta \mathcal{A}_\Delta C_\Delta(k^*) + (1 - F_\Delta \mathcal{A}_\Delta)\right],     
\end{equation}
where $F_\Delta$ is a free parameter representing the number of $\Delta$ resonances contributing to deuteron production divided by the number of all measured deuterons. The parameter $\mathcal{A}_\Delta$ is an arbitrary normalization constant, 
introduced to keep the physically motivated definition of $F_\Delta$ intact. 
The term $C_\Delta(k^*)$ reflects the spectral shape of the $\Delta$ resonance measured and fitted in the $\pi$--p analysis by ALICE (see previous section)~\cite{ALICE:pPi}, transformed to the $\pi$--d system. 
The mass ($M_\Delta=1215~$MeV/$c^2$) and width ($\Gamma_\Delta$) of the $\Delta$ resonance in the present analysis are fixed to the values extracted from the measured $\pi$--p correlations, while the $\Delta$ spectral temperature $T$ is fitted. The width $\Gamma_\Delta$ is $m_\mathrm{T}$ dependent, for the $m_\mathrm{T}$-integrated data shown in Fig.~\ref{fig:pi-dfits} the value is 95~MeV/$c^2$.

The fit to the data is performed in the range $k^*\in(0, 500)$~MeV/$c$, with a systematic variation of $k^*\in(0, 600)$~MeV/$c$. As a systematic check, a 5\% variation in $\lambda_\text{gen}$ is considered, accounting for the uncertainties arising in the determination of secondary contributions and purities due to systematic variations in the particle candidate selection criteria.
Since the parameters $F_\Delta$ and $\mathcal{A}_\Delta$ are maximally correlated, the fit is performed using the effective parameter $F'_\Delta = F_\Delta \mathcal{A}_\Delta$. The parameter $F'_\Delta$ represents the fraction of $\pi$--d pairs in which the pion and at least one of the nucleons within the deuteron originate from a $\Delta$. This can be expressed as 
\begin{equation}\label{eq:Fprim_delta}
    F'_\Delta = \int C_\Delta(k^*)N_\text{mixed}(k^*) dk^* / \int N_\text{mixed}(k^*) dk^*.
\end{equation}
Since the key parameter in this study is $F_\Delta$, establishing a relationship with $F'_\Delta$ is necessary. A straightforward analytical transformation can be derived under the assumption that most recorded collisions containing a reconstructed deuteron include only one. This implies that no additional $\Delta$ signal is introduced in the peak region due to combinatorial effects, and the number of $\Delta$ resonances associated to deuteron production becomes equal to the number of pairs (peak amplitude) linked to a $\Delta$. This results in 
\begin{equation}\label{eq:F_delta}
    F_\Delta \approx \int C_\Delta(k^*)N_\text{mixed}(k^*) dk^* / N_\text{d}, 
\end{equation}
where $N_\text{d}$ is the total number of reconstructed deuterons used in the analysis. Given the fraction of events containing more than one deuteron, the uncertainty associated to Eq.~\ref{eq:F_delta} is estimated to be negligible ($\lesssim0.03\%$). Using Eqs.~\ref{eq:Fprim_delta} and~\ref{eq:F_delta}
\begin{equation}\label{eq:get_F_delta}
    F_\Delta = \frac{\int N_\text{mixed}(k^*)}{N_\text{d}} F'_\Delta = 0.533\pm0.035,
\end{equation}
where both $N_\text{mixed}(k^*)$ and $N_\text{d}$ are measured, while $F'_\Delta$ is extracted from the fit. The quoted uncertainty combines the statistical and systematic errors of the data and the fit.
The fit results for the phase space parameters (Eq.~\ref{Eq. PS Term}) are $p_{\text{T},\Delta} = 985\pm171$~MeV/$c$ and $T=20\pm2$~MeV.

\subsection*{Deuteron and Proton fraction from resonances}

Relating $F_\Delta$ to the probability $P_\Delta$ of producing a single nucleon from a $\Delta$ resonance requires accounting for the reconstruction efficiency. While the efficiency of deuterons cancels out due to the definition of $F_\Delta$, the pion reconstruction efficiency, $\varepsilon_\pi$, must be included. The pion efficiencies are obtained using Monte Carlo simulations produced with PYTHIA 8.2~\cite{Pythia8}, tuned to reproduce pp collisions at 13 TeV, and filtered through the ALICE detector and reconstruction algorithm~\cite{Aamodt:2008zz}.

The following calculations are based purely on combinatorial considerations, without explicitly accounting for the microscopic or kinematical properties of the resonances. The probability of producing exactly one of the two nucleons within the deuteron from a $\Delta$ resonance and detecting the decay pion is $2\varepsilon_\pi P_\Delta(1-P_\Delta)$. The probability of having both nucleons within the deuteron originating from a $\Delta$ resonance and detecting both decay pions is $\varepsilon_\pi^2 P_\Delta^2$, while the probability for the same production scenario when failing to detect one of the pions is $2\varepsilon_\pi(1-\varepsilon_\pi) P_\Delta^2$. 
Note that in the case where both nucleons in the deuteron stem from a $\Delta$, the final state contains a single deuteron and two pions, resulting in two entries in the peak region of the correlation function. Since $F_\Delta$ is defined as the ratio of the number of $\pi$--d pairs to single deuterons, the corresponding term, $\varepsilon_\pi^2 P_\Delta^2$, contributes with twice the number of pairs. 
Taking all these considerations into account and adding all of the terms together leads to 
\begin{equation}\label{eq:FD}
    F_\Delta = 2\varepsilon_\pi P_\Delta.
\end{equation}
Due to the effect of double counting some of the pairs, the result must be transformed into the fraction of (single) deuterons, $f_\Delta$, produced via a $\Delta$ resonance. The definition of $f_\Delta$ is similar to $F_\Delta$, but it removes the double-counting effect by taking the pure term $\varepsilon_\pi^2 P_\Delta^2$ without additional multiplication by 2. This leads to the expression
\begin{equation}\label{eq:fD}
    f_\Delta = 2\varepsilon_\pi P_\Delta\left(1-\frac{\varepsilon_\pi P_\Delta}{2}\right).
\end{equation}
Equations~\ref{eq:FD} and~\ref{eq:fD} account for the pion reconstruction efficiency, $\varepsilon_\pi$, in correcting the single-particle purities. Consequently, $f_\Delta$ is evaluated after applying this efficiency correction. The efficiency-independent result, $f^\mathrm{true}_\Delta$, is obtained by setting $\varepsilon_\pi = 1$ and expressing it in terms of the measured $F_\Delta$
\begin{equation}\label{eq:fD_corr}
    f_\Delta^\mathrm{true} = 2 P_\Delta\left(1-\frac{P_\Delta}{2}\right) = \frac{F_\Delta}{\varepsilon_\pi}\left(1-\frac{F_\Delta}{4\varepsilon_\pi}\right).
\end{equation}
Considering the experimental result $F_\Delta = 0.533\pm0.035$ and a pion reconstruction efficiency of $\varepsilon_\pi = 71.53\pm0.65\%$, evaluated using Monte Carlo simulation and averaged over the transverse momentum range for the pion candidates considered in the analysis, the true fraction is calculated as $f_\Delta^\mathrm{true} = 60.6 \pm 4.1\%$. The uncertainty is propagated by treating the errors of $F_\Delta$ and $\varepsilon_\pi$ as independent.

It should be noted that this value must be considered a lower limit, as it is possible for the pion from the $\Delta$ decay to escape the detector acceptance while the associated nucleus is still reconstructed. This loss of $\Delta$ resonances can only be estimated in a model-dependent manner using Monte Carlo event generators. Using EPOS~3 and PYTHIA~8.3, the loss of $\Delta$ particles due to $\eta$ acceptance effects of the pion is estimated to be $15.5\pm0.5\%$, implying that the value of $F_\Delta$ is underestimated by a similar amount. This number is obtained by calculating the acceptance as a function of \ks folded with the measured \ks distribution of the delta resonance. Re-evaluating $f^\mathrm{true}_\Delta$ using Eq.~\ref{eq:fD_corr} by including in addition the acceptance effect, the result becomes $f^\mathrm{true}_\Delta = 68.7 \pm 4.6\%$.

Similar relations apply to deuterons produced from any resonance. However, the corresponding value is experimentally inaccessible due to the large spectral widths and small individual contributions of the other resonances. By defining the total fraction as $f_\mathrm{R}$ and assuming that the ratio $f_\Delta / f_\mathrm{R} = 0.773\pm0.012$, as predicted by the CSM models, holds for the experimental data, it is possible to extrapolate the acceptance corrected $f^\mathrm{true}_\Delta$ to $f^\mathrm{true}_\mathrm{R} = 88.9 \pm 6.3\%$.

The fraction of deuterons from $\Delta$ resonances was also obtained using the EPOS event generator. For this, all resonances included in EPOS were reweighted using the CSM ThermalFIST with the settings shown in Tab.~\ref{tab:BW_FIST_parameters}. The deuteron formation is simulated using a coalescence afterburner~\cite{Coalescence7} in which the information about the mother particles of the nucleons is conserved. For each nucleon in the deuteron the potential resonance mother is identified and it is checked, whether the nucleon and the corresponding $\pi$ falls within the \pt and $\eta$ acceptance. If at least one nucleon in the deuteron fulfils this criterion, the deuteron is counted as stemming from a resonance. Lastly, the $\eta$-acceptance of $\pi$ is expected to be different in EPOS compared to the measurement, since the spectral shape of the $\Delta$ in the experiment is shifted towards lower \ks values. Losses due to this $\eta$-acceptance increase for increasing \ks, and thus are lower in reality than in the simulation. For this, the acceptance as a function of \ks is averaged using the experimental $\Delta$ spectral shape. The resulting acceptance is 83.6\%, while a similar study with PYTHIA 8.3 gives 85.4\%. Averaging these values and taking the variance as an uncertainty, the Monte-Carlo yields $f_\Delta^\mathrm{true,MC}=53.8\pm3.1\%$, a result compatible with the experimental value of $60.6 \pm 4.1\%$.

While removing the model dependence of this estimation is not feasible, a validation can be performed using $\pi$--p correlation measurements~\cite{ALICE:pPi}. For this purpose, we define the experimental fraction
\begin{equation}
   F_{\Delta\rightarrow \text{p, exp.}} = {Y_\text{exp}(\Delta)}/[{Y_\text{exp}(\text{p})\,\epsilon_\pi}] 
\end{equation}
for the $\pi$--p correlation function, 
where $Y_\text{exp}$ denotes the experimentally measured yields of $\Delta$ and proton candidates. $Y_\text{exp}(\Delta)$ is derived from the spectral shape $R_\Delta(k^*)$ published in Ref.~\cite{ALICE:pPi}, and the proton–pion mixed-event distribution $N_{\text{mixed, p--}\pi^\pm}$ via the relation 
\begin{equation}
    Y_\text{exp}(\Delta) = \int R_\Delta(k^*)N_{\text{mixed, p--}\pi^\pm}\text{d}k^*.
\end{equation} This yields proton fractions of $(14.2 \pm 0.4)\%$ from $\Delta^{++}$ decays and $(5.8 \pm 0.3)\%$ from $\Delta^0$ decays. The performance of the Monte Carlo event generators has been validated by calculating the corresponding model-based fraction $F_{\Delta\rightarrow \text{p, MC}}$, using Thermal-FIST to fix the initial yields of resonances and baryons, followed by PYTHIA or EPOS simulations within the ALICE acceptance. The change in the $\eta$-acceptance of the pion due to the shift of the experimentally observed $\Delta$ spectral shape is accounted for as described above. With this, Thermal-FIST + EPOS predicts that 13.4\% of the protons stem from $\Delta^{++}$ within the experimental acceptance and 4.6\% from $\Delta^0$. The corresponding numbers in Thermal-FIST + PYTHIA are 14.6\% for $\Delta^{++}$ and 5.0\% for $\Delta^0$.

\subsection*{Simulations}
The simulation of the $\pi^+$--d correlation function was performed for three different hypotheses. The first simulation (Fig.~\ref{fig:sketch} (iii)) is done using the EPOS 3 event generator combined with a coalescence afterburner developed in Ref.~\cite{Coalescence7}, and it is able to reproduce the total number of deuterons in the analyzed dataset without any free parameters. The deuterons obtained from this coalescence afterburner are combined with all pions of the desired charge in the same event to create the same event distribution and with a buffer of up to 50 pions from previous events to build the mixed event distribution. The predictions using ThermalFIST use the ThermalFIST
sampler~\cite{Vovchenko2022Jan,TheFIST}, which employs a Cooper-Frye particlization sampling procedure~\cite{Vovchenko2022Jan} and a Blast-Wave parameterization~\cite{Schnedermann1993Nov} tuned to pp collisions at $\sqrt{s}=13$~TeV~\cite{ALICE:BlastWavepp} to obtain positions and momenta of the particles. In the Blast-Wave model, a thermalized medium expands radially with a subsequent instantaneous freeze-out. Its main parameters are the average expansion velocity $\langle\beta\rangle$, its kinetic freeze-out temperature $T_\mathrm{kin}$, and the velocity profile exponent $n$. ThermalFIST can directly produce deuterons without the need for an afterburner, and the same mixed events can be directly constructed in a similar manner as before. The last prediction obtained with ThermalFIST and SMASH uses the particles output by the ThermalFIST sampler in the kinematic region $|\eta|<1$, including deuterons, and feeds it into the hadronic afterburner SMASH~\cite{SMASH}. Inside SMASH, particles rescatter for up to 15 fm/$c$ with a fixed timestep of $\Delta t=0.001$ fm/$c$. The stochastic collision criterion is chosen to enable deuteron production and destruction via $3\leftrightarrow2$ scattering processes such as $\text{p}+\text{n}+\pi\leftrightarrow \text{d}+\pi$. Furthermore, all $2\leftrightarrow2$ processes included in SMASH are enabled. The parameters used in ThermalFIST and the Blast-Wave model are shown in Tab.~\ref{tab:BW_FIST_parameters}

\subsection*{Resonance spectral temperature}

\begin{figure}[!hbt]
    \centering
    \includegraphics[width=0.49\linewidth]{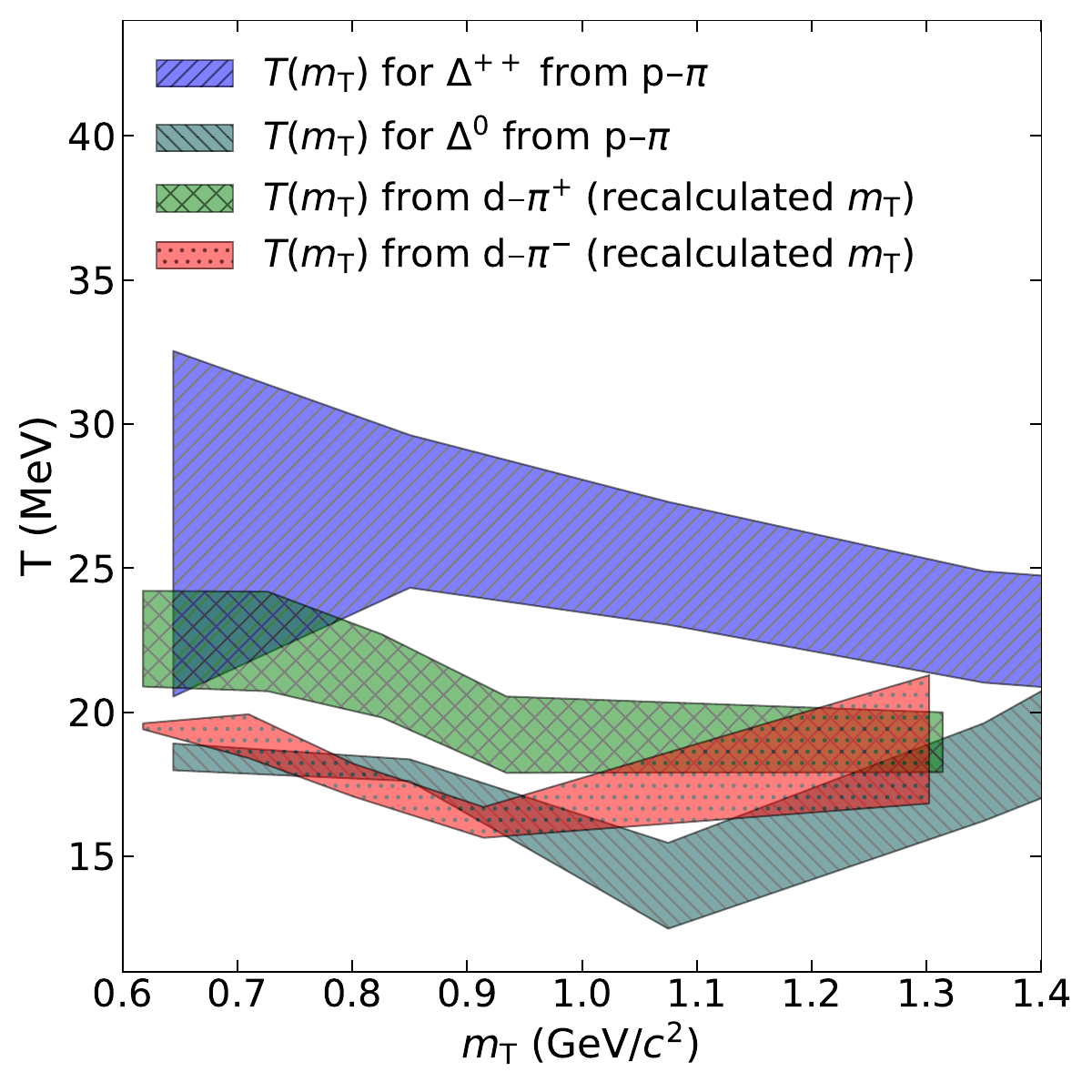}
    \caption{\textbf{Extracted $\Delta$ spectral temperature.} The $\Delta$ spectral temperature is derived from $\pi^{\pm}$--p and \piPMd correlation functions measured in high-multiplicity pp collisions at $\sqrt{s} = 13$~TeV. The bands correspond to the uncertainties obtained by fits to the correlation functions, incorporating systematic uncertainties on the measured data, as well as those arising from variations in the source size and the $\lambda$ parameter for the genuine interaction.}
    \label{fig:temp}
\end{figure}

The modifications of the $\Delta$ spectral shape in the fitting procedure are modeled with the $PS$ term presented in Eq.~\ref{Eq. PS Term}. As discussed above, this function is effectively controlled by the $\Delta$ spectral temperature $T$. The $\Delta$ spectral temperatures for the $\pi^\pm$--d system are shown in Fig.~\ref{fig:temp} as a function of the transverse mass, $m_\text{T}$, of one nucleon in the deuteron and the pion. They are comparable for $\pi^+$--d and $\pi^-$--d but differ from the $\pi^\pm$--p systems discussed in Ref.~\cite{ALICE:pPi}, being lower than the spectral temperature found for $\Delta^{++}$ and higher than that of $\Delta^0$. This aligns qualitatively with the resonance regeneration, $\Delta \leftrightarrow \text{N}\pi$, and rescattering picture~\cite{ Reichert:2019lny, Reichert:2022uha}. For $\pi^+$--p, repulsive strong and Coulomb interactions stop $\Delta^{++}$ regeneration and rescattering earlier, while the attractive $\pi^-$--p interaction allows extended $\Delta^0$ regeneration and rescattering, leading to a lower $\Delta$ spectral temperature.
In $\pi^+$--d, the signal arises from $\Delta^{++}\rightarrow \pi^+$p (with subsequent fusion with a neutron) or $\Delta^{+}\rightarrow \pi^+$n (with later fusion with a proton). Since $\pi^+$--p interactions are repulsive and $\pi^+$--n interactions attractive~\cite{Hoferichter:2015hva}, $\Delta^+$ undergoes longer regeneration cycles than $\Delta^{++}$. This results in a lower spectral temperature than a pure $\Delta^++$ as reflected in the data. Similarly, the $\pi^-$--d system includes contributions from $\Delta^{0}\rightarrow \pi^-$p (attractive) and $\Delta^{-}\rightarrow \pi^-$n (repulsive). The shorter regeneration phase of $\Delta^{-}$ compared to $\Delta^{0}$ yields a higher temperature for $\pi^-$--d than a pure $\pi^-$--p system, again seen in the measurements.

\subsection*{Dibaryon hypothesis and $\pi$--d correlations}
Although no dibaryon states have been confirmed unambiguously (such as a bound N$\Delta$), candidates have been proposed to explain anomalies in data reported by WASA-at-COSY~\cite{PhysRevLett.106.242302}, ELPH~\cite{ELPHDibaryon}, and BGOOD~\cite{BGOODDibaryon}. In particular, the N$\Delta$(2114) candidate was first observed by ELPH with a reported mass of $m_{2B}=2140$~MeV/$c^2$, and later by BGOOD at $m_{2B}=2114$~MeV/$c^2$. The expected imprint of such a state on the $\pi^{\pm}$-d correlation can be evaluated using ThermalFIST. Assuming an extreme upper limit of 100\% for the unknown $\pi$–d branching ratio, as much as $28.5\pm1.5\%$ of deuterons could originate from this decay. Adopting a more realistic branching ratio, estimated from the ratio of elastic to inelastic $\pi$–d cross sections~\cite{PDG22}, reduces this fraction to only $9.5\pm0.5\%$, far below the $60.6\pm4.1\%$ estimated in the present study. Furthermore, if this small contribution is taken as a template, the fit to the data is incompatible with the observed correlation. When the amplitude is left unconstrained, the measured signal remains consistent with at most 1\% of deuterons being produced through such decays. Hence, the experimental data strongly disfavor dibaryon decays as a significant source of the observed signal in the measured $\pi^{\pm}$–d correlation.

\section*{Data Availability Statement}
This manuscript has associated data in a HEPData repository at: \href{https://www.hepdata.net/record/ins2907586}{https://www.hepdata.net/record/ins2907586}.

\section*{Code Availability Statement}
This manuscript has associated code/software in a data repository. The code/software used for the analysis is publicly available on the github repository, at the links \href{https://github.com/alisw/AliRoot}{https://github.com/alisw/AliRoot}, \href{https://github.com/alisw/AliPhysics}{https://github.com/alisw/AliPhysics} and \href{https://github.com/dimihayl/DLM/tree/master/CATS}{https://github.com/dimihayl/DLM/tree/master/CATS}.

\newenvironment{acknowledgement}{\relax}{\relax}
\begin{acknowledgement}
\section*{Acknowledgements}
The ALICE Collaboration is grateful to Elena Bratkovskaya, J\"org Aichelin, Kaijia Sun, Alejandro Kievsky, Michele Viviani, Che-Ming Ko, Stanis\l aw Mr\'owcz\'nski, and Sebastian K\"onig for the fruitful discussions on the pion--deuteron interactions and light (anti)nuclei production.

The ALICE Collaboration would like to thank all its engineers and technicians for their invaluable contributions to the construction of the experiment and the CERN accelerator teams for the outstanding performance of the LHC complex.
The ALICE Collaboration gratefully acknowledges the resources and support provided by all Grid centres and the Worldwide LHC Computing Grid (WLCG) collaboration.
The ALICE Collaboration acknowledges the following funding agencies for their support in building and running the ALICE detector:
A. I. Alikhanyan National Science Laboratory (Yerevan Physics Institute) Foundation (ANSL), State Committee of Science and World Federation of Scientists (WFS), Armenia;
Austrian Academy of Sciences, Austrian Science Fund (FWF): [M 2467-N36] and Nationalstiftung f\"{u}r Forschung, Technologie und Entwicklung, Austria;
Ministry of Communications and High Technologies, National Nuclear Research Center, Azerbaijan;
Conselho Nacional de Desenvolvimento Cient\'{\i}fico e Tecnol\'{o}gico (CNPq), Financiadora de Estudos e Projetos (Finep), Funda\c{c}\~{a}o de Amparo \`{a} Pesquisa do Estado de S\~{a}o Paulo (FAPESP) and Universidade Federal do Rio Grande do Sul (UFRGS), Brazil;
Bulgarian Ministry of Education and Science, within the National Roadmap for Research Infrastructures 2020-2027 (object CERN), Bulgaria;
Ministry of Education of China (MOEC) , Ministry of Science \& Technology of China (MSTC) and National Natural Science Foundation of China (NSFC), China;
Ministry of Science and Education and Croatian Science Foundation, Croatia;
Centro de Aplicaciones Tecnol\'{o}gicas y Desarrollo Nuclear (CEADEN), Cubaenerg\'{\i}a, Cuba;
Ministry of Education, Youth and Sports of the Czech Republic, Czech Republic;
The Danish Council for Independent Research | Natural Sciences, the VILLUM FONDEN and Danish National Research Foundation (DNRF), Denmark;
Helsinki Institute of Physics (HIP), Finland;
Commissariat \`{a} l'Energie Atomique (CEA) and Institut National de Physique Nucl\'{e}aire et de Physique des Particules (IN2P3) and Centre National de la Recherche Scientifique (CNRS), France;
Bundesministerium f\"{u}r Bildung und Forschung (BMBF) and GSI Helmholtzzentrum f\"{u}r Schwerionenforschung GmbH, Germany;
General Secretariat for Research and Technology, Ministry of Education, Research and Religions, Greece;
National Research, Development and Innovation Office, Hungary;
Department of Atomic Energy Government of India (DAE), Department of Science and Technology, Government of India (DST), University Grants Commission, Government of India (UGC) and Council of Scientific and Industrial Research (CSIR), India;
National Research and Innovation Agency - BRIN, Indonesia;
Istituto Nazionale di Fisica Nucleare (INFN), Italy;
Japanese Ministry of Education, Culture, Sports, Science and Technology (MEXT) and Japan Society for the Promotion of Science (JSPS) KAKENHI, Japan;
Consejo Nacional de Ciencia (CONACYT) y Tecnolog\'{i}a, through Fondo de Cooperaci\'{o}n Internacional en Ciencia y Tecnolog\'{i}a (FONCICYT) and Direcci\'{o}n General de Asuntos del Personal Academico (DGAPA), Mexico;
Nederlandse Organisatie voor Wetenschappelijk Onderzoek (NWO), Netherlands;
The Research Council of Norway, Norway;
Pontificia Universidad Cat\'{o}lica del Per\'{u}, Peru;
Ministry of Science and Higher Education, National Science Centre and WUT ID-UB, Poland;
Korea Institute of Science and Technology Information and National Research Foundation of Korea (NRF), Republic of Korea;
Ministry of Education and Scientific Research, Institute of Atomic Physics, Ministry of Research and Innovation and Institute of Atomic Physics and Universitatea Nationala de Stiinta si Tehnologie Politehnica Bucuresti, Romania;
Ministerstvo skolstva, vyskumu, vyvoja a mladeze SR, Slovakia;
National Research Foundation of South Africa, South Africa;
Swedish Research Council (VR) and Knut \& Alice Wallenberg Foundation (KAW), Sweden;
European Organization for Nuclear Research, Switzerland;
Suranaree University of Technology (SUT), National Science and Technology Development Agency (NSTDA) and National Science, Research and Innovation Fund (NSRF via PMU-B B05F650021), Thailand;
Turkish Energy, Nuclear and Mineral Research Agency (TENMAK), Turkey;
National Academy of  Sciences of Ukraine, Ukraine;
Science and Technology Facilities Council (STFC), United Kingdom;
National Science Foundation of the United States of America (NSF) and United States Department of Energy, Office of Nuclear Physics (DOE NP), United States of America.
In addition, individual groups or members have received support from:
Czech Science Foundation (grant no. 23-07499S), Czech Republic;
FORTE project, reg.\ no.\ CZ.02.01.01/00/22\_008/0004632, Czech Republic, co-funded by the European Union, Czech Republic;
European Research Council (grant no. 950692), European Union;
Deutsche Forschungs Gemeinschaft (DFG, German Research Foundation) ``Neutrinos and Dark Matter in Astro- and Particle Physics'' (grant no. SFB 1258), Germany;
ICSC - National Research Center for High Performance Computing, Big Data and Quantum Computing and FAIR - Future Artificial Intelligence Research, funded by the NextGenerationEU program (Italy);
National Recovery and Resilience Plan of the Republic of Bulgaria, project SUMMIT BG-RRP-2.004-0008-C01, funded by the NextGenerationEU program (Bulgaria).
\end{acknowledgement}

\section*{Extended data}
\label{sec:ExtendedData}

\begin{table}[ht]
\begin{center}
    \caption{\textbf{Parameters used in the measured correlation function.} The values of the measured average transverse mass $\langle m_{\mathrm{T}}\rangle$, extracted source sizes $r_\text{core}$, $r_\text{eff}$, and  $\lambda$ parameter which serves as the weights for the contribution of the genuine \piPd and \piMd pairs in the measured correlation function.}
    \begin{tabular}{| c | c |c | c | c| c |}
    \hline 
        \mt interval & \mt range (GeV/$c^2$)& $\langle m_{\mathrm{T}}\rangle$ (GeV/$c^2$) & $r_\text{core}^{\small\piPMd}$ (fm) & $r_\text{eff}^{\small\piPMd}$ (fm)&$\lambda_{\text{gen}}^{\piPMd}$ \\
         \hline 
         integrated & $1.03-2.24$& 1.27 & $1.08\pm0.04$ & $1.51\pm0.12$ & $81.6$\%\\
           \hline
    \end{tabular}
    \label{tab:TabmValuesSourceLambda}
\end{center}
\end{table}

\begin{table}[!hbt]
    \centering    
    \caption{\textbf{Parameters used in the $\pi^+$--d correlation function predictions for thermal production.} Parameters on the left side are used for the Blast-Wave parameterization, and parameters on the right are used for the ThermalFIST yields.}
    \label{tab:BW_FIST_parameters}
    \begin{tabular}{c||c|c|c|c||c|c}
         Parameter& Value & Unit &&Parameter&Value & Unit \\
         \hline\hline
         $\langle\beta\rangle$ & 0.51 & --&&$T_\mathrm{c}$ & 0.165 & GeV\\
         $T_\mathrm{kin}$ & 0.16 & GeV&&$\mu_B$ & 0 & --\\
         $n$ & 1.4 & --&&$\gamma_S$ & 0.85 & --\\
         $\eta_\mathrm{max}$ & 1.5 & --&&$\mathrm{d} V/\mathrm{d} y$ & 75 & fm$^{3}$\\
         $R_0$ & 1.8 & fm&&$V_c$ & 3 & $\mathrm{d} V/\mathrm{d} y$\\
         Ref. & \cite{ALICE:BlastWavepp}&--&Ref. & \cite{vanillaCSM}&--
    \end{tabular}

\end{table}

\clearpage

\section{The ALICE Collaboration}
\label{app:collab}
\begin{flushleft} 
\small

S.~Acharya\,\orcidlink{0000-0002-9213-5329}\,$^{\rm 50}$, 
A.~Agarwal$^{\rm 133}$, 
G.~Aglieri Rinella\,\orcidlink{0000-0002-9611-3696}\,$^{\rm 32}$, 
L.~Aglietta\,\orcidlink{0009-0003-0763-6802}\,$^{\rm 24}$, 
M.~Agnello\,\orcidlink{0000-0002-0760-5075}\,$^{\rm 29}$, 
N.~Agrawal\,\orcidlink{0000-0003-0348-9836}\,$^{\rm 25}$, 
Z.~Ahammed\,\orcidlink{0000-0001-5241-7412}\,$^{\rm 133}$, 
S.~Ahmad\,\orcidlink{0000-0003-0497-5705}\,$^{\rm 15}$, 
S.U.~Ahn\,\orcidlink{0000-0001-8847-489X}\,$^{\rm 71}$, 
I.~Ahuja\,\orcidlink{0000-0002-4417-1392}\,$^{\rm 36}$
% ...

\section*{Affiliation Notes}

$^{\rm I}$ Deceased\\
$^{\rm II}$ Also at: Max-Planck-Institut fur Physik, Munich, Germany\\
$^{\rm III}$ Also at: Italian National Agency for New Technologies, Energy and Sustainable Economic Development (ENEA), Bologna, Italy\\
$^{\rm IV}$ Also at: Instituto de Fisica da Universidade de Sao Paulo\\
$^{\rm V}$ Also at: Dipartimento DET del Politecnico di Torino, Turin, Italy\\
$^{\rm VI}$ Also at: Department of Applied Physics, Aligarh Muslim University, Aligarh, India\\
$^{\rm VII}$ Also at: Institute of Theoretical Physics, University of Wroclaw, Poland\\
$^{\rm VIII}$ Also at: Facultad de Ciencias, Universidad Nacional Aut\'{o}noma de M\'{e}xico, Mexico City, Mexico\\

\section*{Collaboration Institutes}

$^{1}$ A.I. Alikhanyan National Science Laboratory (Yerevan Physics Institute) Foundation, Yerevan, Armenia\\
$^{2}$ AGH University of Krakow, Cracow, Poland\\
$^{3}$ Bogolyubov Institute for Theoretical Physics, National Academy of Sciences of Ukraine, Kiev, Ukraine\\
$^{4}$ Bose Institute, Department of Physics  and Centre for Astroparticle Physics and Space Science (CAPSS), Kolkata, India\\
$^{5}$ California Polytechnic State University, San Luis Obispo, California, United States\\
$^{6}$ Central China Normal University, Wuhan, China\\
$^{7}$ Centro de Aplicaciones Tecnol\'{o}gicas y Desarrollo Nuclear (CEADEN), Havana, Cuba\\
$^{8}$ Centro de Investigaci\'{o}n y de Estudios Avanzados (CINVESTAV), Mexico City and M\'{e}rida, Mexico\\
$^{9}$ Chicago State University, Chicago, Illinois, United States\\
$^{10}$ China Nuclear Data Center, China Institute of Atomic Energy, Beijing, China\\
$^{11}$ China University of Geosciences, Wuhan, China\\
$^{12}$ Chungbuk National University, Cheongju, Republic of Korea\\
$^{13}$ Comenius University Bratislava, Faculty of Mathematics, Physics and Informatics, Bratislava, Slovak Republic\\
$^{14}$ Creighton University, Omaha, Nebraska, United States\\
$^{15}$ Department of Physics, Aligarh Muslim University, Aligarh, India\\
$^{16}$ Department of Physics, Pusan National University, Pusan, Republic of Korea\\
$^{17}$ Department of Physics, Sejong University, Seoul, Republic of Korea\\
$^{18}$ Department of Physics, University of California, Berkeley, California, United States\\
$^{19}$ Department of Physics, University of Oslo, Oslo, Norway\\
$^{20}$ Department of Physics and Technology, University of Bergen, Bergen, Norway\\
$^{21}$ Dipartimento di Fisica, Universit\`{a} di Pavia, Pavia, Italy\\
$^{22}$ Dipartimento di Fisica dell'Universit\`{a} and Sezione INFN, Cagliari, Italy\\
$^{23}$ Dipartimento di Fisica dell'Universit\`{a} and Sezione INFN, Trieste, Italy\\
$^{24}$ Dipartimento di Fisica dell'Universit\`{a} and Sezione INFN, Turin, Italy\\
$^{25}$ Dipartimento di Fisica e Astronomia dell'Universit\`{a} and Sezione INFN, Bologna, Italy\\
$^{26}$ Dipartimento di Fisica e Astronomia dell'Universit\`{a} and Sezione INFN, Catania, Italy\\
$^{27}$ Dipartimento di Fisica e Astronomia dell'Universit\`{a} and Sezione INFN, Padova, Italy\\
$^{28}$ Dipartimento di Fisica `E.R.~Caianiello' dell'Universit\`{a} and Gruppo Collegato INFN, Salerno, Italy\\
$^{29}$ Dipartimento DISAT del Politecnico and Sezione INFN, Turin, Italy\\
$^{30}$ Dipartimento di Scienze MIFT, Universit\`{a} di Messina, Messina, Italy\\
$^{31}$ Dipartimento Interateneo di Fisica `M.~Merlin' and Sezione INFN, Bari, Italy\\
$^{32}$ European Organization for Nuclear Research (CERN), Geneva, Switzerland\\
$^{33}$ Faculty of Electrical Engineering, Mechanical Engineering and Naval Architecture, University of Split, Split, Croatia\\
$^{34}$ Faculty of Nuclear Sciences and Physical Engineering, Czech Technical University in Prague, Prague, Czech Republic\\
$^{35}$ Faculty of Physics, Sofia University, Sofia, Bulgaria\\
$^{36}$ Faculty of Science, P.J.~\v{S}af\'{a}rik University, Ko\v{s}ice, Slovak Republic\\
$^{37}$ Faculty of Technology, Environmental and Social Sciences, Bergen, Norway\\
$^{38}$ Frankfurt Institute for Advanced Studies, Johann Wolfgang Goethe-Universit\"{a}t Frankfurt, Frankfurt, Germany\\
$^{39}$ Fudan University, Shanghai, China\\
$^{40}$ Gangneung-Wonju National University, Gangneung, Republic of Korea\\
$^{41}$ Gauhati University, Department of Physics, Guwahati, India\\
$^{42}$ Helmholtz-Institut f\"{u}r Strahlen- und Kernphysik, Rheinische Friedrich-Wilhelms-Universit\"{a}t Bonn, Bonn, Germany\\
$^{43}$ Helsinki Institute of Physics (HIP), Helsinki, Finland\\
$^{44}$ High Energy Physics Group,  Universidad Aut\'{o}noma de Puebla, Puebla, Mexico\\
$^{45}$ Horia Hulubei National Institute of Physics and Nuclear Engineering, Bucharest, Romania\\
$^{46}$ HUN-REN Wigner Research Centre for Physics, Budapest, Hungary\\
$^{47}$ Indian Institute of Technology Bombay (IIT), Mumbai, India\\
$^{48}$ Indian Institute of Technology Indore, Indore, India\\
$^{49}$ INFN, Laboratori Nazionali di Frascati, Frascati, Italy\\
$^{50}$ INFN, Sezione di Bari, Bari, Italy\\
$^{51}$ INFN, Sezione di Bologna, Bologna, Italy\\
$^{52}$ INFN, Sezione di Cagliari, Cagliari, Italy\\
$^{53}$ INFN, Sezione di Catania, Catania, Italy\\
$^{54}$ INFN, Sezione di Padova, Padova, Italy\\
$^{55}$ INFN, Sezione di Pavia, Pavia, Italy\\
$^{56}$ INFN, Sezione di Torino, Turin, Italy\\
$^{57}$ INFN, Sezione di Trieste, Trieste, Italy\\
$^{58}$ Inha University, Incheon, Republic of Korea\\
$^{59}$ Institute for Gravitational and Subatomic Physics (GRASP), Utrecht University/Nikhef, Utrecht, Netherlands\\
$^{60}$ Institute of Experimental Physics, Slovak Academy of Sciences, Ko\v{s}ice, Slovak Republic\\
$^{61}$ Institute of Physics, Homi Bhabha National Institute, Bhubaneswar, India\\
$^{62}$ Institute of Physics of the Czech Academy of Sciences, Prague, Czech Republic\\
$^{63}$ Institute of Space Science (ISS), Bucharest, Romania\\
$^{64}$ Institut f\"{u}r Kernphysik, Johann Wolfgang Goethe-Universit\"{a}t Frankfurt, Frankfurt, Germany\\
$^{65}$ Instituto de Ciencias Nucleares, Universidad Nacional Aut\'{o}noma de M\'{e}xico, Mexico City, Mexico\\
$^{66}$ Instituto de F\'{i}sica, Universidade Federal do Rio Grande do Sul (UFRGS), Porto Alegre, Brazil\\
$^{67}$ Instituto de F\'{\i}sica, Universidad Nacional Aut\'{o}noma de M\'{e}xico, Mexico City, Mexico\\
$^{68}$ iThemba LABS, National Research Foundation, Somerset West, South Africa\\
$^{69}$ Jeonbuk National University, Jeonju, Republic of Korea\\
$^{70}$ Johann-Wolfgang-Goethe Universit\"{a}t Frankfurt Institut f\"{u}r Informatik, Fachbereich Informatik und Mathematik, Frankfurt, Germany\\
$^{71}$ Korea Institute of Science and Technology Information, Daejeon, Republic of Korea\\
$^{72}$ Laboratoire de Physique Subatomique et de Cosmologie, Universit\'{e} Grenoble-Alpes, CNRS-IN2P3, Grenoble, France\\
$^{73}$ Lawrence Berkeley National Laboratory, Berkeley, California, United States\\
$^{74}$ Lund University Department of Physics, Division of Particle Physics, Lund, Sweden\\
$^{75}$ Nagasaki Institute of Applied Science, Nagasaki, Japan\\
$^{76}$ Nara Women{'}s University (NWU), Nara, Japan\\
$^{77}$ National and Kapodistrian University of Athens, School of Science, Department of Physics , Athens, Greece\\
$^{78}$ National Centre for Nuclear Research, Warsaw, Poland\\
$^{79}$ National Institute of Science Education and Research, Homi Bhabha National Institute, Jatni, India\\
$^{80}$ National Nuclear Research Center, Baku, Azerbaijan\\
$^{81}$ National Research and Innovation Agency - BRIN, Jakarta, Indonesia\\
$^{82}$ Niels Bohr Institute, University of Copenhagen, Copenhagen, Denmark\\
$^{83}$ Nikhef, National institute for subatomic physics, Amsterdam, Netherlands\\
$^{84}$ Nuclear Physics Group, STFC Daresbury Laboratory, Daresbury, United Kingdom\\
$^{85}$ Nuclear Physics Institute of the Czech Academy of Sciences, Husinec-\v{R}e\v{z}, Czech Republic\\
$^{86}$ Oak Ridge National Laboratory, Oak Ridge, Tennessee, United States\\
$^{87}$ Ohio State University, Columbus, Ohio, United States\\
$^{88}$ Physics department, Faculty of science, University of Zagreb, Zagreb, Croatia\\
$^{89}$ Physics Department, Panjab University, Chandigarh, India\\
$^{90}$ Physics Department, University of Jammu, Jammu, India\\
$^{91}$ Physics Program and International Institute for Sustainability with Knotted Chiral Meta Matter (WPI-SKCM$^{2}$), Hiroshima University, Hiroshima, Japan\\
$^{92}$ Physikalisches Institut, Eberhard-Karls-Universit\"{a}t T\"{u}bingen, T\"{u}bingen, Germany\\
$^{93}$ Physikalisches Institut, Ruprecht-Karls-Universit\"{a}t Heidelberg, Heidelberg, Germany\\
$^{94}$ Physik Department, Technische Universit\"{a}t M\"{u}nchen, Munich, Germany\\
$^{95}$ Politecnico di Bari and Sezione INFN, Bari, Italy\\
$^{96}$ Research Division and ExtreMe Matter Institute EMMI, GSI Helmholtzzentrum f\"ur Schwerionenforschung GmbH, Darmstadt, Germany\\
$^{97}$ Saga University, Saga, Japan\\
$^{98}$ Saha Institute of Nuclear Physics, Homi Bhabha National Institute, Kolkata, India\\
$^{99}$ School of Physics and Astronomy, University of Birmingham, Birmingham, United Kingdom\\
$^{100}$ Secci\'{o}n F\'{\i}sica, Departamento de Ciencias, Pontificia Universidad Cat\'{o}lica del Per\'{u}, Lima, Peru\\
$^{101}$ Stefan Meyer Institut f\"{u}r Subatomare Physik (SMI), Vienna, Austria\\
$^{102}$ SUBATECH, IMT Atlantique, Nantes Universit\'{e}, CNRS-IN2P3, Nantes, France\\
$^{103}$ Sungkyunkwan University, Suwon City, Republic of Korea\\
$^{104}$ Suranaree University of Technology, Nakhon Ratchasima, Thailand\\
$^{105}$ Technical University of Ko\v{s}ice, Ko\v{s}ice, Slovak Republic\\
$^{106}$ The Henryk Niewodniczanski Institute of Nuclear Physics, Polish Academy of Sciences, Cracow, Poland\\
$^{107}$ The University of Texas at Austin, Austin, Texas, United States\\
$^{108}$ Universidad Aut\'{o}noma de Sinaloa, Culiac\'{a}n, Mexico\\
$^{109}$ Universidade de S\~{a}o Paulo (USP), S\~{a}o Paulo, Brazil\\
$^{110}$ Universidade Estadual de Campinas (UNICAMP), Campinas, Brazil\\
$^{111}$ Universidade Federal do ABC, Santo Andre, Brazil\\
$^{112}$ Universitatea Nationala de Stiinta si Tehnologie Politehnica Bucuresti, Bucharest, Romania\\
$^{113}$ University of Derby, Derby, United Kingdom\\
$^{114}$ University of Houston, Houston, Texas, United States\\
$^{115}$ University of Jyv\"{a}skyl\"{a}, Jyv\"{a}skyl\"{a}, Finland\\
$^{116}$ University of Kansas, Lawrence, Kansas, United States\\
$^{117}$ University of Liverpool, Liverpool, United Kingdom\\
$^{118}$ University of Science and Technology of China, Hefei, China\\
$^{119}$ University of South-Eastern Norway, Kongsberg, Norway\\
$^{120}$ University of Tennessee, Knoxville, Tennessee, United States\\
$^{121}$ University of the Witwatersrand, Johannesburg, South Africa\\
$^{122}$ University of Tokyo, Tokyo, Japan\\
$^{123}$ University of Tsukuba, Tsukuba, Japan\\
$^{124}$ Universit\"{a}t M\"{u}nster, Institut f\"{u}r Kernphysik, M\"{u}nster, Germany\\
$^{125}$ Universit\'{e} Clermont Auvergne, CNRS/IN2P3, LPC, Clermont-Ferrand, France\\
$^{126}$ Universit\'{e} de Lyon, CNRS/IN2P3, Institut de Physique des 2 Infinis de Lyon, Lyon, France\\
$^{127}$ Universit\'{e} de Strasbourg, CNRS, IPHC UMR 7178, F-67000 Strasbourg, France, Strasbourg, France\\
$^{128}$ Universit\'{e} Paris-Saclay, Centre d'Etudes de Saclay (CEA), IRFU, D\'{e}partment de Physique Nucl\'{e}aire (DPhN), Saclay, France\\
$^{129}$ Universit\'{e}  Paris-Saclay, CNRS/IN2P3, IJCLab, Orsay, France\\
$^{130}$ Universit\`{a} degli Studi di Foggia, Foggia, Italy\\
$^{131}$ Universit\`{a} del Piemonte Orientale, Vercelli, Italy\\
$^{132}$ Universit\`{a} di Brescia, Brescia, Italy\\
$^{133}$ Variable Energy Cyclotron Centre, Homi Bhabha National Institute, Kolkata, India\\
$^{134}$ Warsaw University of Technology, Warsaw, Poland\\
$^{135}$ Wayne State University, Detroit, Michigan, United States\\
$^{136}$ Yale University, New Haven, Connecticut, United States\\
$^{137}$ Yildiz Technical University, Istanbul, Turkey\\
$^{138}$ Yonsei University, Seoul, Republic of Korea\\
$^{139}$ Affiliated with an institute formerly covered by a cooperation agreement with CERN\\
$^{140}$ Affiliated with an international laboratory covered by a cooperation agreement with CERN.\\

\end{flushleft} 
  
\end{document}